\documentclass[preprint,ejp]{revtex4-1}
\usepackage{amsmath}
\usepackage{amsfonts}

\begin{document}

\title{A large class of bound-state solutions of the Schr\"{o}dinger
equation via Laplace transform of the confluent hypergeometric equation\footnote{To appear in Journal of Mathematical Chemistry}}
\author{P. H. F. Nogueira}
\email{pedrofusconogueira@gmail.com}
\author{A. S. de Castro}
\email{castro@pq.cnpq.br}\address{Universidade Estadual Paulista, Campus de Guaratinguet\'{a}, Departamento de F\'{\i}sica e
Qu\'{\i}mica, \protect{12516-410} Guaratinguet\'{a}, SP,  Brazil}
\author{D. R. M. Pimentel}
\email{douglas.roberto.fis@gmail.com}\address{Universidade de S\~ao Paulo, Instituto de F\'{\i}sica, Departamento de F\'{\i}sica Experimental, 05508-090
S\~ao Paulo, SP, Brazil}

\begin{abstract}
It is shown that analytically soluble bound states of the Schr\"{o}dinger
equation for a large class of systems relevant to atomic and molecular
physics can be obtained by means of the Laplace transform of the confluent
hypergeometric equation. It is also shown that all closed-form
eigenfunctions are expressed in terms of generalized Laguerre polynomials.
The generalized Morse potential is used as an illustration.
\end{abstract}

\maketitle

\section{Introduction}

Some exactly soluble systems with importance in atomic and molecular physics
have been approached in the literature on quantum mechanics with a myriad of
methods. Among such systems is the Morse potential $a(e^{-\alpha
x}-2e^{-2\alpha x})$ \cite{b1mor}-\cite{b1don3}, the pseudoharmonic
potential $a\left( x/b-b/x\right) ^{2}$ \cite{b1gol}-\cite{b1haa}, \cite%
{b2wei}-\cite{b2tez}, and the Kratzer-Fues potential $a\left(
b^{2}/x^{2}-2b/x\right) $ and its modified version $a\left(
b^{2}/x^{2}-b/x\right) $ \cite{b1gol}-\cite{b1haa}, \cite{b1flu}, \cite%
{b2ikh2}-\cite{b3mol}. More general exactly soluble systems have also been
appreciated: the generalized Morse potential $Ae^{-\alpha x}+Be^{-2\alpha x}$
\cite{b2tez}, \cite{b4bar}-\cite{b4ard1}, the singular harmonic oscillator $%
Ax^{2}+Bx^{-2}$ \cite{b1lan}-\cite{b1haa}, \cite{b1don3}, \cite{b4bag}, \cite%
{b5con}-\cite{b5asc}, the singular Coulomb potential \thinspace $%
Ax^{-1}+Bx^{-2}$ \cite{con}-\cite{b1haa}, \cite{b4bag}, \cite{b5con}, \cite%
{b5don1}, \cite{b5ikh}-\cite{b5ard2}, \cite{b6hal}-\cite{b6asc}, and some
ring-shaped potentials (see, e.g. \cite{b1don3} and references therein).

The integral transform methods have proven to be useful and powerful for
solving ordinary differential equations because they can convert the
original equation into a simpler differential equation or into an algebraic
equation. The Laplace transform method applied to quantum mechanics was used
by Schr\"{o}dinger into the discussion of radial eigenfunction of the
hydrogen atom \cite{sch}, and later Englefield approached the
three-dimensional Schr\"{o}dinger equation with diverse spherically
symmetric potentials \cite{eng}. More than twenty years later the hydrogen
atom was reexamined with the Laplace transform method \cite{swa}. Recently,
some interest has been revived in searching bound-state solutions of the Schr%
\"{o}dinger equation via Laplace transform method. For some years now
one-dimensional problems with the $1/x$ \cite{ran}, Morse \cite{che1},
generalized Morse \cite{ard1}, Dirac delta \cite{asc} and harmonic
oscillator \cite{pim} potentials, three-dimensional problems with the
singular harmonic oscillator, the singular Coulomb \cite{ard2}, some
ring-shaped \cite{ard3} potentials, and the $D$-dimensional harmonic
oscillator \cite{che2}, have been solved for the Laplace transform. With
fulcrum on the relation mapping the behaviour of the eigenfunction near
infinity and the Laplace transform near isolated singular points, Englefield
\cite{eng} found the spectrum of three-dimensional problems by imposing that
the radial eigenfunction vanishes at the origin. Englefield's recipe, spiced
up with the relation mapping the behaviour of the eigenfunction near the
origin and its corresponding transform near infinity, was used in Ref. \cite%
{pim} in order to obtain the complete set of bound-state solutions for the
one-dimensional harmonic oscillator without using the closed-form solution
for the Laplace transform. Furthermore, the class of problems was enlarged
to include eigenfunctions satisfying homogeneous Neumann conditions at the
origin.

In the present paper, the spiced Englefield's recipe is followed with
attention restricted to systems that after factorizing the behaviour at the
neighbourhood of special points, the second-order differential equation for
the eigenfunction can be reduced to the confluent hypergeometric equation.
Then is shown that all well-behaved eigenfunctions for that class of systems
are expressed in terms of generalized Laguerre polynomials. Exactly solvable
problems in this category include all the potentials mentioned in the first
paragraph and the exactly soluble generalized Morse potential is used as an
illustration.

\section{Laplace transform and a few of its properties}

Let us begin with a brief description of the Laplace transform and a few of
its properties \cite{doe}. The Laplace transform of a function $\Phi $ is
defined by
\begin{equation}
F(s)=\mathcal{L}\left\{ \Phi \right\} =\int_{0}^{\infty }d\xi \,e^{-s\xi
}\Phi \left( \xi \right) .  \label{l1}
\end{equation}%
If there is some positive constant $\sigma $ such that $\Phi $ does not
increase faster than $e^{\sigma \xi }$ for sufficiently large $\xi $ then $%
\Phi $ is said to be of exponential order $\sigma $. In this case, the
integral in Equation (\ref{l1}) may exist for \textrm{Re}$\,s>\sigma $.
Nevertheless, the Laplace transform may fail to exist because of a
sufficiently strong singularity in the function $\Phi $ as $\xi \rightarrow 0
$. In particular, $\xi ^{\lambda }$ is of exponential order arbitrary and
\begin{equation}
\mathcal{L}\left\{ \frac{\xi ^{\lambda }}{\Gamma \left( \lambda +1\right) }%
\right\} =\frac{1}{s^{\lambda +1}},\quad \mathrm{Re}\,\lambda >-1,\quad
\mathrm{Re}\,s>0,  \label{177}
\end{equation}%
where $\Gamma $ is the gamma function. Derivative properties involving the
Laplace transform are convenient for solving differential equations. In this
paper we shall use the following properties:
\begin{eqnarray}
\mathcal{L}\left\{ \frac{d\Phi }{d\xi }\right\}  &=&sF\left( s\right)
-\left. \Phi \right\vert _{\xi =0}  \notag \\
&&  \notag \\
\mathcal{L}\left\{ \frac{d^{2}\Phi }{d\xi ^{2}}\right\}  &=&s^{2}F\left(
s\right) -s\left. \Phi \right\vert _{\xi =0}-\left. \frac{d\Phi }{d\xi }%
\right\vert _{\xi =0}  \label{13} \\
&&  \notag \\
\mathcal{L}\left\{ \xi \Phi \right\}  &=&-\frac{dF\left( s\right) }{ds}.
\notag
\end{eqnarray}%
More than this, we shall use a pair of relations mapping limiting forms. If
near an isolated singular point $s_{0}$ the Laplace transform behaves as%
\begin{equation}
F\left( s\right) \underset{s\rightarrow s_{0}}{\sim }\frac{1}{\left(
s-s_{0}\right) ^{\nu }},\quad \nu >0,  \label{14}
\end{equation}%
then
\begin{equation}
\Phi \left( \xi \right) \underset{\xi \rightarrow \infty }{\sim }\frac{1}{%
\Gamma \left( \nu \right) }\,\xi ^{\nu -1}\,e^{s_{0}\xi }.  \label{15}
\end{equation}%
On the other hand,%
\begin{equation}
\underset{s\rightarrow \infty }{\lim }sF\left( s\right) =\Phi \left(
0\right) ,  \label{16}
\end{equation}%
an result known as initial value theorem.

\section{ The generalized Morse potential}

The time-independent Schr\"{o}dinger equation is an eigenvalue equation for
the characteristic pair $(E,\psi )$ with $E\in
\mathbb{R}
$. For a particle of mass $m$ embedded in the generalized Morse potential it
reads
\begin{equation}
\frac{d^{2}\psi \left( x\right) }{dx^{2}}+\frac{2m}{\hbar ^{2}}\left(
E-V_{1}e^{-\alpha x}-V_{2}e^{-2\alpha x}\right) \psi \left( x\right) =0,
\label{sch}
\end{equation}%
where $\hbar $ is Planck's constant, $\alpha >0$, and $\int_{-\infty
}^{+\infty }dx\,|\psi |^{2}=1$ for bound states. The substitution%
\begin{equation}
\xi =\frac{2\sqrt{2mV_{2}}\,e^{-\alpha x}}{\hbar \alpha }  \label{xi}
\end{equation}%
and the definitions%
\begin{equation}
S=\frac{\sqrt{-2mE}}{\hbar \alpha },\quad a=\frac{mV_{1}}{\hbar \alpha \sqrt{%
2mV_{2}}}+S+\frac{1}{2}  \label{k}
\end{equation}%
convert Eq. (\ref{sch}) into%
\begin{equation}
\frac{d^{2}\psi \left( \xi \right) }{d\xi ^{2}}+\frac{1}{\xi }\frac{d\psi
\left( \xi \right) }{d\xi }+\left( -\frac{1}{4}+\frac{S-a+1/2}{\xi }-\frac{%
S^{2}}{\xi ^{2}}\right) \psi \left( \xi \right) =0,  \label{whi}
\end{equation}%
whose solutions have asymptotic limits expressed as $\psi \left( \xi \right)
\underset{|\xi |\rightarrow 0}{\rightarrow }\xi ^{\pm S}$ and $\psi \left(
\xi \right) \underset{|\xi |\rightarrow \infty }{\rightarrow }e^{\pm \xi /2}$%
. On account of the normalization condition, $\int_{0}^{\infty }d|\xi
|\,|\psi \left( \xi \right) |^{2}/|\xi |=\alpha $, one has that $\psi $
behaves like $\xi ^{S}$ as $|\xi |\rightarrow 0$ and like $e^{-\xi /2}$ as $%
|\xi |\rightarrow \infty $ with $\xi \in
\mathbb{R}
$ ($V_{2}>0$) and $S>0$ ($E<0$). The substitution%
\begin{equation}
\psi \left( \xi \right) =e^{-\xi /2}\xi ^{S}\Phi \left( \xi \right)
\label{17}
\end{equation}%
transforms Eq. (\ref{whi}) into
\begin{equation}
\xi \,\frac{d^{\,2}\Phi \left( \xi \right) }{d\xi ^{2}}+(b-\xi )\,\frac{%
d\Phi \left( \xi \right) }{d\xi }-a\,\Phi \left( \xi \right) =0,  \label{kum}
\end{equation}%
with $b=2S+1>1$. Eq. (\ref{kum}) is the standard form of the confluent
hypergeometric equation \cite{leb}. Notice that $\Phi $ is a nonzero
constant at the origin and tends to infinity no more rapidly than $\exp
\left( \sigma _{1}\xi ^{\sigma _{2}}\right) $, with $\sigma _{2}<1$ and
arbitrary $\sigma _{1}$, for sufficiently large $\xi $. This occurs because $%
\sigma _{1}\xi ^{\sigma _{2}}-\xi /2\rightarrow -\xi /2$ as $\xi \rightarrow
\infty $. The regular behaviour of $\Phi $ at the origin plus its behaviour
for large $\xi $ ensure the existence of its Laplace transform. In this case
$\Phi $ is of exponential order arbitrary and consequently its Laplace
transform exists for $\mathrm{Re}\,s>0$.

\section{Laplace transform of the confluent hypergeometric equation}

Using the derivative properties of the Laplace transforms given by (\ref{13}%
), the confluent hypergeometric equation is mapped onto%
\begin{equation}
s\left( s-1\right) \frac{dF\left( s\right) }{ds}+\left[ \left( 2-b\right)
s+a-1\right] F\left( s\right) =\left( 1-b\right) \Phi \left( 0\right) ,\quad
\mathrm{Re}\,s>0.  \label{e1}
\end{equation}%
Note that this first-order differential equation has regular (nonessential)
singularities at $s=0$ and $s=1$. Therefore, $F\left( s\right) $ is either
analytical, or possess a pole or branch point, at the regular singular point
(Fuchs theorem). The relation connecting the behaviour of $F$ near an
isolated singular point and the behaviour of $\Phi $ for large $\xi $
dictates that $\Phi $ behaves like $\xi ^{\nu -1}$ or $\xi ^{\nu -1}e^{\xi }$%
, depending on where the isolated singularity of $F$ is, whether near $s=0$
or $s=1$, respectively. Due to the asymptotic behaviour prescribed for $\Phi
$ at the end of the previous section, one sees that $F$ behaves like $%
s^{-\nu }$ as $s\rightarrow 0$, and (\ref{e1}) enforces%
\begin{equation}
\nu =1-a.  \label{ast1}
\end{equation}%
On the other hand, using the initial value theorem one sees that $F$ behaves
like $\Phi \left( 0\right) /s$ as $s\rightarrow \infty $. Thus, we seek a
particular solution of (\ref{e1}), regular at $s=1$, in the form of a
polynomial in inverse powers of $s$:
\begin{equation}
F\left( s\right) =\sum_{j=0}^{n}c_{j}s^{j-\nu }=\frac{c_{0}}{s^{\nu }}%
+\cdots +\frac{c_{n}}{s},  \label{e2}
\end{equation}%
with $c_{0}\neq 0$ and%
\begin{equation}
\nu =n+1,  \label{ast2}
\end{equation}
in such a way that $s=0$ is a pole of order $n+1$ and $F\left( s\right) $ is
the principal part of a finite Laurent series with residue $c_{n}=\Phi
\left( 0\right) $ at $s=0$. Comparing (\ref{ast1}) with (\ref{ast2}), one
sees that

\begin{equation}
a=-n.  \label{qc}
\end{equation}%
Substituting (\ref{e2}) into (\ref{e1}) one obtains the following two-term
recursive relation for the coefficients%
\begin{equation}
c_{j+1}=c_{j}\frac{1+j-n-b}{j+1},\quad j\geq 0.
\end{equation}%
Inspection and induction yields%
\begin{equation}
c_{j}=c_{0}\frac{\left( -1\right) ^{j}}{j!}\frac{\Gamma \left( n+b\right) }{%
\Gamma (n+b-j)},\quad 0\leq j\leq n.  \label{c2}
\end{equation}%
This means that%
\begin{equation}
F\left( s\right) =c_{0}\sum_{j=0}^{n}\frac{\left( -1\right) ^{j}}{j!}\frac{%
\Gamma \left( n+b\right) }{\Gamma \left( n+b-j\right) }s^{j-n-1}.  \label{c3}
\end{equation}%
Using (\ref{177}), the termwise inverse transformation of (\ref{c3}) leads
to the polynomial solution for $\Phi $:
\begin{equation}
\Phi \left( \xi \right) =c_{0}\sum_{j=0}^{n}\frac{\left( -1\right) ^{j}}{j!}%
\frac{\Gamma \left( n+b\right) }{\Gamma \left( n+b-j\right) \left(
n-j\right) !}\xi ^{n-j}.  \label{c4}
\end{equation}%
Then, using Leibniz's formula for the generalized Laguerre polynomials \cite%
{leb}%
\begin{equation}
L_{n}^{\left( b-1\right) }\left( \xi \right) =\sum_{j=0}^{n}\frac{\Gamma
\left( n+b\right) }{\Gamma \left( j+b\right) }\frac{\left( -\xi \right) ^{j}%
}{j!\left( n-j\right) !},\quad b>0,
\end{equation}%
one obtains%
\begin{equation}
\Phi _{n}\left( \xi \right) =c_{0}\left( -1\right) ^{n}L_{n}^{\left(
b-1\right) }\left( \xi \right) .  \label{c5}
\end{equation}%
Actually, condition (\ref{qc}) transforms the confluent hypergeometric
equation into generalized Laguerre's equation in such a way that the
succeeding process involving the inversion of the Laplace transform is not
surprising.

For systems whose eigenfunctions can be expressed in terms of a particular
solution of the confluent hypergeometric equation, Eqs. (\ref{qc}) and (\ref%
{c5}) summarize all we need to determine the complete set of bound-state
solutions.

\section{Bound states in a generalized Morse potential}

We now turn our attention to the generalized Morse potential. Substitution
of (\ref{qc}) into (\ref{k}) leads to the quantization condition%
\begin{equation}
n+S+\frac{1}{2}=-\frac{mV_{1}}{\hbar \alpha \sqrt{2mV_{2}}}.  \label{QC}
\end{equation}%
Hence, $V_{1}<0$ so that the generalized Morse potential is able to hold
bound states only if it has a well structure ($V_{1}<0$ and $V_{2}>0$).
Furthermore, because $E<0$ one gets%
\begin{equation}
n<\frac{m|V_{1}|}{\hbar \alpha \sqrt{2mV_{2}}}-\frac{1}{2}.
\end{equation}%
This restriction on $n$ limits the number of allowed states and requires $%
m|V_{1}|/\left( \hbar \alpha \sqrt{2mV_{2}}\right) >1/2$ to make the
existence of a bound state possible. Finally, we use the quantization
condition (\ref{QC}) to write%
\begin{equation}
E_{n}=-\frac{V_{1}^{2}}{4V_{2}}\left[ 1-\frac{\hbar \alpha \sqrt{2mV_{2}}}{%
m|V_{1}|}\left( n+\frac{1}{2}\right) \right] ^{2}.
\end{equation}

With the generalized Laguerre polynomial standardized as \cite{leb}
\begin{equation}
L_{n}^{\left( \mu \right) }\left( z\right) =\sum\limits_{j=0}^{n}\frac{%
\Gamma \left( n+\mu +1\right) }{\Gamma \left( j+\mu +1\right) }\frac{\left(
-z\right) ^{j}}{j!\left( n-j\right) !}  \label{leb}
\end{equation}%
and the integral \cite{gr}%
\begin{equation}
\int\limits_{0}^{\infty }dz\,e^{-z}z^{\gamma -1}L_{n}^{\left( \mu \right)
}\left( z\right) =\frac{\Gamma \left( \gamma \right) \Gamma \left( 1+\mu
+n-\gamma \right) }{n!\,\Gamma \left( 1+\mu -\gamma \right) },\quad \text{%
Re\thinspace }\gamma >0,
\end{equation}%
one can show that
\begin{equation}
\int\limits_{0}^{\infty }dz\,e^{-z}z^{\mu -1}\left[ L_{n}^{\left( \mu
\right) }\left( z\right) \right] ^{2}=\frac{\Gamma \left( \mu +n+1\right) }{%
\mu \,n!},\quad \text{Re}\text{ }\mu >0.  \label{nieto}
\end{equation}%
Therefore, the normalization condition yields the normalized eigenfunction
(firstly obtained in Ref. \cite{b1nie} for $V_{2}=-2V_{1}$):
\begin{equation}
\psi _{n}\left( \xi \right) =\sqrt{\frac{2\alpha S\,n!}{\Gamma \left(
2S+n+1\right) }}\,\xi ^{S}e^{-\xi /2}L_{n}^{\left( 2S\right) }\left( \xi
\right) .  \label{psinorm}
\end{equation}

\section{Concluding remarks}

Bessel's equation as well as differential equations with linear coefficients
can be mapped onto simpler homogeneous first-order differential equations
for the Laplace transform when the original functions are subject to special
boundary conditions at the origin (see, e.g. \cite{doe}). Following the
spiced Englefield's recipe, we have shown that the bound-state solutions of
the Schr\"{o}dinger equation whose eigenfunctions are expressed in terms of
particular solutions of the confluent hypergeometric equation, including the
large class of systems with potentials relevant to atomic and molecular
physics such as the generalized Morse, singular harmonic oscillator, and
singular Coulomb potentials, beyond the ring-shaped potentials approached in
Ref. \cite{ard3}, can be obtained by using the Laplace transform of the
confluent hypergeometric equation. In those cases, the Laplace transform
maps the confluent hypergeometric equation onto a nonhomogeneous first-order
differential equation. The source of nonhomogeneity is just the particular
solution of the confluent hypergeometric equation at the origin but this
fact does not represent a less favourable position because it is related
asymptotically to the residue of the Laplace transform at the origin via the
initial value theorem. It is worthwhile recall that the eigenfunction for
the generalized Morse potential does not have values prescribed at the
origin. The spiced Englefield's recipe allows searching for a particular
solution of the transformed equation with a well-defined singularity and a
well-defined asymptotic behaviour, and such as presented in this paper the
exact-closed form of the Laplace transform does not have relevance to
determinate the complete set of bound-state solutions.

\begin{acknowledgments}
This work was supported in part by means of funds provided by  FAPESP, CNPq and CAPES.
\end{acknowledgments}

\end{document}